\documentclass[12pt]{article}
\usepackage{amssymb}
\usepackage{amsmath}
\usepackage{graphicx}

\newcommand{\be}{\begin{equation}}
\newcommand{\ee}{\end{equation}}
\newcommand{\bea}{\begin{eqnarray}}
\newcommand{\eea}{\end{eqnarray}}

\begin{document}

\title{\textbf{Stable Exact Solutions in Cosmological Models with Two Scalar Fields}}

\author{I.~Ya.~Aref'eva$^1$\footnote{arefeva@mi.ras.ru}, \
N.~V.~Bulatov$^2$\footnote{nick\_bulatov@mail.ru}, \
S.~Yu.~Vernov$^3$\footnote{svernov@theory.sinp.msu.ru} \\[2.7mm]
${}^1$\small{Steklov Mathematical Institute, Russian Academy of Sciences,}\\
\small{Gubkina str. 8, 119991, Moscow, Russia}\\
${}^2$\small{Department of Quantum Statistics and Field Theory, Faculty
of Physics,}\\ \small{Moscow State University, Leninskie
Gory 1, 119991, Moscow, Russia}\\
${}^3$\small{Skobeltsyn Institute of Nuclear Physics, Moscow State University},\\
\small{Leninskie Gory 1, 119991, Moscow, Russia}\\
}

\date{ }

\maketitle

\begin{abstract}
 The stability of isotropic cosmological solutions for two-field models in
the Bianchi I metric is considered. We prove that  the sufficient
conditions  for the Lyapunov stability in the
Friedmann--Robertson--Walker metric provide the stability with respect
to  anisotropic perturbations in the Bianchi I metric and with respect
to the cold dark matter energy density fluctuations. Sufficient
conditions for the Lyapunov stability of the isotropic fixed points of
the system of the Einstein equations have been found. We use the
superpotential method to construct stable kink-type solutions and
obtain sufficient conditions on the superpotential for the Lyapunov
stability of the corresponding exact solutions. We analyze the
stability of isotropic kink-type solutions for string field theory
inspired cosmological models.
\end{abstract}

\section{Introduction}
 To specify different components of the cosmic fluid one
typically uses a phenomenological relation $p=w\varrho$ between the
pressure (Lagrangian density) $p$ and the energy density $\varrho$
corresponding to each component of the fluid. The function $w$ is
called the state parameter. Contemporary
observations~\cite{cosmo-obser} give strong support that in the
Universe, the uniformly distributed cosmic fluid with negative
pressure, the so-called dark energy, currently dominates with a state
parameter value approximately equal to $-1$:
\begin{equation*}
w_{DE} = -1 \pm 0.2.
\end{equation*}
Strong restrictions on the anisotropy were found using
observations~\cite{Barrow,Bernui:2005pz}, and it was also shown that
the Universe is spatially flat at large distances. It has been shown
in~\cite{9,Nesseris,ZhangGui} (see also
reviews~\cite{Quinmodrev1,Quinmodrev2} and references therein), that
the recent analysis of the observation data indicates that the time
dependent state parameter gives a better fit than $w_{DE}\equiv -1$,
corresponding to the cosmological constant. It, in particular, gives
reasons for the interest in models with $w_{DE} < -1$. The field theory
with $w_{DE} < -1$ is also interesting as a possible solution of the
cosmological singularity problem~\cite{Hawking-Ellis,cyclic,GV}.

The standard way to obtain an evolving state parameter is to add scalar
fields into a cosmological model.  Two-field models with the crossing
of the cosmo\-logical constant barrier $w_{DE}=-1$ are known as quintom
models and include one phantom scalar field and one ordinary scalar
field. Quintom models are being actively studied at present
time~\cite{Quinmodrev1,Quinmodrev2,Guo2004,AKV2,Lazkoz,Leon,QuintomModels}.
The cosmological models with $w_{DE}<-1$ violate the null energy
condition (NEC), this violation is generally related to the phantom
fields appearing. The standard quantization of these models leads to
instability, which is physically unacceptable.
In~\cite{AKV2}, the theory with $w_{DE}<-1$ was interpreted as an
approximation in the framework of the fundamental theory. Because the
fundamental theory must be stable and must admit quantization, this
instability can be considered an artefact of the approximation.

We note the problems of quantum instability that arise in effective
theories and are related to high order derivatives~\cite{AV-NEC,RAS}.
In~\cite{SW,Creminelli-eff} the instability problem has been reduced to
the problem of such choose of the effective theory parameters that the
instability turns out to be essential only at times that are not
described in the framework of the effective theory approximation. In
the mathematical language it means that the terms with higher-order
derivatives can be treated as corrections essential only at small
energies below the physical cut-off. This approach implies the
possibility  to construct a UV completion of the theory and to consider
these effective theories physically acceptable with the presumption
that an effective theory admits immersion into a fundamental theory.

In the Friedmann--Robertson--Walker (FRW) metric the NEC violating
models can have classically stable solutions cosmology. In particular,
there are classically stable solutions for ghost models with minimal
coupling to  gravity. Moreover, there exists an attractor behavior in a
class of the phantom cosmological
models~\cite{phantom-attractor,AKVCDM,Lazkoz} (attractor solutions for
inhomogeneous cosmological models were considered
in~\cite{Starobinsky}).

The stability of isotropic solutions in the Bianchi
models~\cite{Bianchi,Ellis98,Ellis} (see also~\cite{DSC}) has been
considered in inflationary models (see~\cite{Germani,Koivisto} and
references therein for details of anisotropic slow-roll
inflation). Assuming that the energy conditions are satisfied, it
has been proved that all initially expanding Bianchi models except
type IX approach the de Sitter space-time~\cite{Wald} (see
also~\cite{MossSahni,Wainwright,Kitada,Rendall04}).   The Wald
theorem~\cite{Wald} shows that for space-time of Bianchi types
I--VIII with a positive cosmological constant and matter
satisfying the dominant and strong energy conditions, solutions
which exist globally in the future have certain asymptotic
properties at $t\rightarrow\infty$.

The standard way to analyse the stability of solutions for Friedmann
equations in quintom models uses the change of
variables~\cite{Guo2004,Lazkoz,Leon} (see also~\cite{Quinmodrev2}). In
the case of exponential potentials such a transformation is useful,
because it transforms the class of nontrivial solutions to fixed points
of the new system of equations~\cite{Guo2004}. We show that the
stability conditions for an arbitrary potential that were found
in~\cite{Lazkoz} can be obtained without introducing any new variables.

Studying the stability of solutions in the FRW metric, we specify a
form of fluctuations. It is interesting to know whether these
(isotropic) solutions are stable under the deformation of the FRW
metric to an anisotropic one, for example, to the Bianchi I metric. In
comparison with  general fluctuations we can get an explicit form of
fluctuations in the Bianchi I metric, which can probably clarify some
nontrivial issues of theories with the NEC violation.

In this paper we consider the stability of isotropic solutions in the
Bianchi I metric. Interpreting the solutions of the Friedmann equation
as isotropic solutions in the Bianchi I metric, we include anisotropic
perturbations in our consideration. The stability analysis is
essentially simplified by a suitable choice of
variables~\cite{Ellis98,DSC,Pereira}. In this paper we show that for an
arbitrary potential the stability conditions, obtained
in~\cite{Lazkoz}, are sufficient for stability not only in the FRW
metric, but also in the Bianchi I metric. We also analyse the stability
with respect to small fluctuations of the initial value of the cold
dark matter energy density. We consider quintom cosmological models, as
well as models with two scalar (or two phantom scalar) fields.

The stability of a continuous solution tending to a fixed point implies
  the stability of this fixed point. Using the Lyapunov theorem \cite{Lyapunov,Pontryagin} we find
conditions under which the fixed point and the corresponding kink (or
lump) solution are stable. For one field models the sufficient
conditions for stability of isotropic solutions, which tend to fixed
points, have been obtained in~\cite{ABJV09}.

In~\cite{AKV2,Vernov06} the superpotential method has been used to
construct quintom models with exact solutions. In this paper we get the
stability conditions in terms of the superpotential and use the
superpotential for construction of two-field models with stable exact
solutions. We also verify the stability of solutions, obtained in the
string field theory (SFT) inspired quintom models~\cite{AKV2,Vernov06}.

The paper is organized as follows. In Section 2 we consider two-field
model with an arbitrary potential and we find the sufficient stability
conditions in the FRW and Bianchi I metrics. In Section 3 we remind the
superpotential method and obtain conditions on the superpotential which
are sufficient for the stability of exact solutions.  In Section 4 we
check the stability of kink-type solutions in a few SFT inspired
cosmological models. In Section~5 we summarize the results and make a
conclusion. In Appendix we present the stability conditions on the
superpotential in the case of one-field cosmological models.

\section{Sufficient stability conditions}

\subsection{The Einstein equations in the Bianchi I metric}

We consider a two-field cosmological model with the following action
\begin{equation}
 S=\int d^4x \sqrt{-g}\left[\frac{R}{16\pi G_N}-
 \frac{C_1}{2}g^{\mu\nu}\partial_{\mu}\phi\partial_{\nu}\phi-
 \frac{C_2}{2}g^{\mu\nu}\partial_{\mu}\xi\partial_{\nu}\xi
-V(\phi,\xi)\right], \label{action_2}
\end{equation}
where the potential $V(\phi,\xi)$ is a twice continuously
differentiable function, which can include the cosmological constant
$\Lambda$, $G_N$ is the Newtonian gravitational constant ($8\pi
G_N=1/M_P^2$, where $M_P$ is the Planck mass). Each of the fields
$\phi$ and $\xi$ is either scalar or phantom scalar fields in
dependence on signs of the constants $C_1$ and $C_2$.

Let us consider the Bianchi I metric
\begin{equation} \label{Bianchi}
{ds}^{2}={}-{dt}^2+a_1^2(t)dx_1^2+a_2^2(t)dx_2^2+a_3^2(t)dx_3^2.
\end{equation}
It is convenient to express $a_i$ in terms of new variables $a$ and
$\beta_i$ (we use the notation in~\cite{Pereira}):
\begin{equation}
a_i(t)= a(t) e^{\beta_i(t)}.
\end{equation}

Imposing the constraint
\begin{equation}
\label{restr1} \beta_1+\beta_2+\beta_3=0,
\end{equation}
one has the following relations
\begin{equation}
a(t)=(a_1(t)a_2(t)a_3(t))^{1/3},
\end{equation}
\begin{equation}\label{Hi}
H_i\equiv \dot a_i/a_i= H+\dot\beta_i, \qquad\mbox{and}\qquad H\equiv
\dot a/a=\frac{1}{3}(H_1+H_2+H_3),
\end{equation}
where the dot denotes the time derivative. To obtain (\ref{Hi}) we have
used the following  consequence of (\ref{restr1}):
\begin{equation}
\label{restr2} \dot\beta_1+\dot\beta_2+\dot\beta_3=0.
\end{equation}

Note that $\beta_i$ are not components of a vector and, therefore,
are not subjected to the Einstein summation rule. In the case of
the FRW metric all $\beta_i$ are equal to zero and $H$ is the
Hubble parameter. Following~\cite{Ellis98,Pereira} (see
also~\cite{DSC}) we introduce the shear
\begin{equation}
\sigma^2\equiv \dot\beta_1^2+\dot\beta_2^2+\dot\beta_3^2.
\end{equation}

The Einstein equations have the following form:
\begin{equation}
\label{a2} 3H^2-\frac{1}{2}\sigma^2=8\pi G_N\varrho,
\end{equation}
\begin{equation}
\label{trequ} 2\dot H+3H^2+\frac{1}{2}\sigma^2={}-8\pi G_Np,
\end{equation}
\begin{equation}
\label{d2} \dot\phi=\psi,\qquad
\dot{\psi}={}-3H\psi-\frac{1}{C_1}\frac{\partial V}{\partial\phi},
\end{equation}
\begin{equation}
\label{e2} \dot\xi=\zeta,\qquad
\dot{\zeta}={}-3H\zeta-\frac{1}{C_2}\frac{\partial
V}{\partial\xi},
\end{equation}
where
\begin{equation}
\label{varrho_pressure} \varrho=
\frac{C_1}{2}\dot{\phi}^2+\frac{C_2}{2}\dot{\xi}^2+V(\phi,\xi),\qquad
    p=\frac{C_1}{2}\dot{\phi}^2+\frac{C_2}{2}\dot{\xi}^2-V(\phi,\xi).
\end{equation}

For $\beta_i$ and $\sigma^2$ we obtain the following equations
\begin{equation}
\ddot\beta_i={}-3H\dot \beta_i, \label{equbeta}
\end{equation}
\begin{equation}
\label{equvartheta}
\frac{d}{dt}\left(\sigma^2\right)={}-6H\sigma^2.
\end{equation}

Functions $H(t)$, $\sigma^2(t)$, $\phi(t)$, $\xi(t)$, and $\rho(t)$ can
be obtained from equations (\ref{a2})--(\ref{e2}) and
(\ref{equvartheta}). If $H(t)$ is known then $\beta_i$ can be trivially
obtained from (\ref{equbeta}). We show in the next subsection that
functions $H(t)$, $\dot\beta_i(t)$ and $\sigma^2(t)$ are very suitable
to analyse the stability of isotropic solutions in the Bianchi I
metric.

\subsection{Sufficient conditions for the Lyapunov stability of a fixed
point}

Summing equations (\ref{a2}) and (\ref{trequ}) we obtain the following
system
\begin{equation}
\label{SYSTEM2}
\begin{array}{l}
\displaystyle \dot H={}-3H^2+8\pi G_NV(\phi,\xi),\\
\displaystyle \dot\phi=\psi,\\
\displaystyle \dot{\psi}={}-3H\psi-\frac{1}{C_1}\frac{\partial V}{\partial\phi},\\
\displaystyle \dot\xi=\zeta,\\
\displaystyle \dot{\zeta}={}-3H\zeta-\frac{1}{C_2}\frac{\partial
V}{\partial\xi}.
\end{array}
\end{equation}

Let the fields $\phi$ and $\xi$ tend to finite limits as $t\rightarrow
+\infty$. System (\ref{SYSTEM2}) has a fixed point
$y_f=(H_f,\phi_f,\psi_f,\xi_f,\zeta_f)$ if and only if
\begin{eqnarray}
 H_f^2&=&\displaystyle \frac{8\pi G_N}{3} V(\phi_{f},\xi_{f}),
\label{l}\\
\psi_{f}&=&0,\label{h}
\\
\zeta_{f}&=&0, \label{i}
\\
V'_{\phi}& \equiv&\frac{\partial V}{\partial\phi}(\phi_{f},\xi_{f})=0,
\label{j}
\\
V'_{\xi}&\equiv&  \frac{\partial V}{\partial\xi}(\phi_{f},\xi_{f})= 0,
\label{k}
\end{eqnarray}

All fixed points $y_f$ correspond to $\psi_{f}=0$ and $\zeta_{f}=0$. We
denote the fixed point $y_f=(H_f,\phi_f,\psi_f,0,0)$ as
$y_f=(H_f,\phi_f,\psi_f)$. To analyse the stability of $y_f$ we study
the stability of this fixed point for the corresponding linearized
system of equations and use the Lyapunov
theorem~\cite{Lyapunov,Pontryagin}. In the neighborhood of $y_f$ we
have
\begin{eqnarray}
H(t)&=&H_f + \varepsilon h_1(t) +{\cal O}(\varepsilon^2),
\label{m}
\\
\phi(t)&=&\phi_{f}+\varepsilon  \phi_1(t)+{\cal O}(\varepsilon^2),
\label{n}
\\
\psi(t)&=&\varepsilon \psi_1(t)+{\cal O}(\varepsilon^2), \label{o}
\\
\xi(t)&=&\xi_{f}+\varepsilon \xi_1(t)+{\cal O}(\varepsilon^2),
\label{p}
\\
\zeta(t)&=&\varepsilon \zeta_1(t)+{\cal O}(\varepsilon^2)
\label{q},
\end{eqnarray}
where $\varepsilon$ is a small parameter.

To first order in $\varepsilon$ we obtain the following system of
equations
\begin{eqnarray}
\displaystyle \dot h_1(t)&=&\displaystyle {}-6H_fh_1(t), \label{r}
\\
\displaystyle \dot\phi_1(t)&=&\displaystyle \psi_1(t), \label{s}
\\
\displaystyle \dot\psi_1(t)&=&\displaystyle {}-3H_f\psi_1(t)-
\frac{1}{C_1}\left(V''_{\phi\phi} \phi_1(t) +V''_{\phi\xi}
\xi_1(t)\right), \label{t}
\\
\displaystyle \dot\xi_1(t)&=&\displaystyle \zeta_1(t), \label{u}
\\
\displaystyle \dot\zeta_1(t)&=&\displaystyle {}-3H_f\zeta_1(t)-
\frac{1}{C_2}\left(V''_{\xi\phi} \phi_1(t) +V''_{\xi\xi}
\xi_1(t)\right), \label{v}
\end{eqnarray}
where
\begin{equation*}
 V''_{\phi\phi}\equiv \frac{\partial ^2V}{\partial \phi ^2}
(\phi_{f},\xi_{f}),\qquad V''_{\xi\xi}\equiv \frac{\partial
^2V}{\partial \xi ^2} (\phi_{f},\xi_{f}),\qquad V''_{\phi\xi}\equiv
\frac{\partial ^2V}{\partial \phi \partial \xi} (\phi_{f},\xi_{f}).
\end{equation*}

Equation (\ref{r}) has the following solution
\begin{equation}
 h(t)=b_0e^{-6H_ft}, \label{w}
\end{equation}
where $b_0$ is a constant. For asymptotic stability of the fixed point
$y_f$ the function $h(t)$ should tend to zero at $t\rightarrow\infty$,
therefore, the asymptotic stability requires that the condition $H_f>0$
be satisfied.

The system of four first order equations (\ref{s})--(\ref{v}) can be
written as the following system of two second order equations
\begin{eqnarray}
&\displaystyle \ddot{\phi}_1(t)+3H_f\dot{\phi}_1(t)+
\frac{1}{C_1}\left(V''_{\phi\phi}\phi_1(t) +V''_{\phi\xi}
 \xi_1(t)\right)=0, \label{x}
\\
&\displaystyle \ddot{\xi}_1(t)+3H_f\dot{\xi}_1(t)+
\frac{1}{C_2}\left( V''_{\phi\xi}\phi_1(t) +V''_{\xi\xi}\xi_1(t)
\right)=0 \label{y}.
\end{eqnarray}

In the case when $V_{\phi\xi}''=0$ the system of equations
(\ref{x})--(\ref{y}) becomes a system of two independent second order
equations. The general solution of this system  is as follows:

\begin{itemize}

\item $\phi_1(t)=\tilde{D}_1e^{-\left(\frac{3}{2}H_f -
\frac{1}{2}\sqrt{9H_f^2-4\frac{V_{\phi\phi}''}{C_1}}\right)t}+\tilde{D}_2e^{-\left(\frac{3}{2}H_f
+ \frac{1}{2}\sqrt{9H_f^2-4\frac{V_{\phi\phi}''}{C_1}}\right)t}$ \
if $9H_f^2\neq 4\frac{V_{\phi\phi}''}{C_1}$,

\item
$\phi_1(t)=\left(\tilde{D}_1+\tilde{D}_2t\right)e^{-\frac{3}{2}H_ft}$ \
if $9H_f^2= 4\frac{V_{\phi\phi}''}{C_1}$,

\end{itemize}
\begin{itemize}

\item $\xi_1(t)=\tilde{D}_3e^{-\left(\frac{3}{2}H_f -
\frac{1}{2}\sqrt{9H_f^2-4\frac{V_{\xi\xi}''}{C_2}}\right)t}+\tilde{D}_4e^{-\left(\frac{3}{2}H_f
+ \frac{1}{2}\sqrt{9H_f^2-4\frac{V_{\xi\xi}''}{C_2}}\right)t}$ \ if
$9H_f^2\neq 4\frac{V_{\xi\xi}''}{C_2}$,

\item
$\xi_1(t)=\left(\tilde{D}_3+\tilde{D}_4t\right)e^{-\frac{3}{2}H_ft}$ \  if
$9H_f^2= 4\frac{V_{\xi\xi}''}{C_2}$,

\end{itemize}
where $\tilde{D}_1$, $\tilde{D_2}$, $\tilde{D_3}$ and $\tilde{D_4}$ are
arbitrary constants.

For the asymptotical stability of the considered fixed point the
functions $\phi_1(t)$ and $\xi_1(t)$ must converge to $0$ at
$t\rightarrow\infty$. We obtain that at $V_{\phi\xi}''=0$ the
sufficient conditions for the asymptotical stability are
\begin{equation}
H_f>0,\qquad \frac{V_{\xi\xi}''}{C_2} >0, \qquad
\frac{V_{\phi\phi}''}{C_1}>0. \label{V2fCondVFKVFK}
\end{equation}

Let us consider the case $V_{\phi\xi}''\neq 0$.
\begin{enumerate}
\item For
\begin{equation*}
{V_{\phi\xi}''}^2\neq\frac{C_1C_2}{16}\left(9H_f^2-\frac{4V_{\xi\xi}''}{C_2}\right)
\left(9H_f^2-\frac{4V_{\phi\phi}''}{C_1}\right)\quad\mbox{and}\quad
{V_{\phi\xi}''}^2 \neq{} -
\frac{C_1C_2}{4}\left(\frac{V_{\xi\xi}''}{C_2}-\frac{V_{\phi\phi}''}{C_1}\right)^2
\end{equation*}
the general solution of (\ref{x})--(\ref{y}) is
\begin{equation}
\begin{array} {c}
\displaystyle{\phi_1(t)}=\tilde{D}_1e^{-(\frac{3}{2}H_f +
\frac{1}{2}\sqrt{\Delta_1-2\sqrt{\Delta_2}})t}+\tilde{D_2}e^{-(\frac{3}{2}H_f
- \frac{1}{2}\sqrt{\Delta_1-2\sqrt{\Delta_2}})t}+{}\\
\displaystyle{}+\tilde{D_3}e^{-(\frac{3}{2}H_f+\frac{1}{2}\sqrt{\Delta_1+2\sqrt{\Delta_2}})t}+
\tilde{D_4}e^{-(\frac{3}{2}H_f-\frac{1}{2}\sqrt{\Delta_1+2\sqrt{\Delta_2}})t},
\end{array}
\label{z}
\end{equation}
\begin{equation}
\begin{array} {c}
\displaystyle{\xi_1(t)}=\frac{C_1V_{\xi\xi}''-C_2V_{\phi\phi}''+
\sqrt{(C_1V_{\xi\xi}''-C_2V_{\phi\phi}'')^2+4C_1C_2{V_{\phi\xi}''}^2}}{2C_2V_{\phi\xi}''}\times{} \\
\displaystyle{}\times\left(\tilde{D}_1e^{-(\frac{3}{2}H_f +
\frac{1}{2}\sqrt{\Delta_1-2\sqrt{\Delta_2}})t}+
\tilde{D_2}e^{-(\frac{3}{2}H_f -
\frac{1}{2}\sqrt{\Delta_1-2\sqrt{\Delta_2}})t}\right)+{} \\
\displaystyle{}+\frac{C_1V_{\xi\xi}''-C_2V_{\phi\phi}''-
\sqrt{(C_1V_{\xi\xi}''-C_2V_{\phi\phi}'')^2+4C_1C_2{V_{\phi\xi}''}^2}}{2C_2V_{\phi\xi}''}\times
\\{}\times \left(\tilde{D_3}e^{-(\frac{3}{2}H_f+\frac{1}{2}\sqrt{\Delta_1+2\sqrt{\Delta_2}})t}
+\tilde{D_4}e^{-(\frac{3}{2}H_f-\frac{1}{2}\sqrt{\Delta_1+2\sqrt{\Delta_2}})t}\right),
\end{array}
\label{aa}
\end{equation}
here and further $\tilde{D}_1$, $\tilde{D_2}$, $\tilde{D_3}$, and
$\tilde{D_4}$ are arbitrary constants and

\begin{equation}
\Delta_1=9H_f^2-2\frac{V_{\xi\xi}''}{C_2}-2\frac{V_{\phi\phi}''}{C_1},
\qquad \Delta_2=\left(\frac{V_{\xi\xi}''}{C_2}
-\frac{V_{\phi\phi}''}{C_1}\right)^2+4\frac{{V_{\phi\xi}''}^2}{C_1C_2}.
\label{abb}
\end{equation}

For asymptotical stability of the considered fixed point these
functions must converge to $0$. As we can see both functions are just
linear combinations of exponents to some degrees. To satisfy this
condition all these degrees must be negative. It is easy to obtain that
the sufficient conditions for asymptotic stability are
\begin{equation}
H_f>0, \qquad \frac{V_{\xi\xi}''}{C_2}+\frac{V_{\phi\phi}''}{C_1}>0,
\qquad
\frac{V_{\xi\xi}''V_{\phi\phi}''}{C_1C_2}>\frac{{V_{\phi\xi}''}^2}{C_1C_2}.
\label{V2fCond}
\end{equation}
\item In the case
\begin{equation*}
{V_{\phi\xi}''}^2=\frac{C_1C_2}{16}\left(9H_f^2-\frac{4V_{\xi\xi}''}{C_2}\right)
\left(9H_f^2-\frac{4V_{\phi\phi}''}{C_1}\right)\quad \mbox{and}\quad
{V_{\phi\xi}''}^2\neq -
\frac{C_1C_2}{4}\left(\frac{V_{\xi\xi}''}{C_2}-\frac{V_{\phi\phi}''}{C_1}\right)^2
\end{equation*}
the inequality $\Delta_1\neq 0$ is valid and the general solution of
(\ref{x})--(\ref{y}) is

\begin{equation}
\begin{array} {c}
\displaystyle{\phi_1(t)}=\tilde{D}_1e^{-\frac{3}{2}H_ft} +
\displaystyle\tilde{D_2}e^{-\frac{3}{2}H_ft}t+
\displaystyle\tilde{D_3}e^{(-\frac{3}{2}H_f+\frac{\sqrt{2}}{2}\sqrt{\Delta_1
})t}+
\displaystyle\tilde{D_4}e^{(-\frac{3}{2}H_f-\frac{\sqrt{2}}{2}\sqrt{\Delta_1
})t},
\end{array}
\label{ab}\
\end{equation}
\begin{equation}
\begin{array} {c}
\displaystyle{\xi_1(t)}=\frac{C_1}{4V_{\phi\xi}''}

\left(\left(9H_f^2-\frac{4V_{\phi\phi}''}{C_1}\right)\left(\tilde{D}_1e^{-\frac{3}{2}H_f t}
+
\displaystyle\tilde{D_2}e^{-\frac{3}{2}H_ft}t\right)\right.-{}
\\[2.7mm]
{}-\displaystyle\left.\left(9H_f^2-\frac{4V_{\xi\xi}''}{C_2}\right)
\left(\tilde{D_3}e^{(-\frac{3}{2}H_f+\frac{\sqrt{2}}{2}\sqrt{\Delta_1
})t}+
\displaystyle\tilde{D_4}e^{(-\frac{3}{2}H_f-\frac{\sqrt{2}}{2}\sqrt{\Delta_1
})t}\right)\right).
\end{array}
\label{ac}
\end{equation}

It is easy to show that in this case the sufficient conditions for
asymptotic stability of the considered fixed point coincide with
(\ref{V2fCondVFKVFK}).

\item In the case $V_{\phi\xi}^2={} -
\frac{C_1C_2}{4}\left(\frac{V_{\xi\xi}}{C_2}-\frac{V_{\phi\phi}}{C_1}\right)^2$,
in other words $\Delta_2=0$, and \\
$V_{\phi\xi}^2\neq\frac{C_1C_2}{16}\left(9H_f^2-\frac{4V_{\xi\xi}}{C_2}\right)
\left(9H_f^2-\frac{4V_{\phi\phi}}{C_1}\right)$ the inequality
$\Delta_1\neq 0$ holds and therefore
\begin{equation*}
\phi_1(t)=\left(\tilde{D}_1+\tilde{D_3}t\right)e^{-(3H_f -
\sqrt{\Delta_1})t/2}+\left(\tilde{D_2}+\tilde{D_4}t\right)e^{-(H_f +
\sqrt{\Delta_1})t/2}, \label{ad}
\end{equation*}
\begin{equation*}
\begin{array} {c}
\displaystyle{\xi_1(t)}=\frac{C_1}{\sqrt{-C_1C_2}}
\left\{\left(\tilde{D}_1+\left(1-\frac{C_1\sqrt{\Delta_1}}{V_{\phi\xi}''}\right)
\tilde{D}_3 t\right)e^{-(3H_f -
\sqrt{\Delta_1})t/2}\right.+{} \\[2.7mm]
{}+\displaystyle\left.
\left(\tilde{D}_2+\left(1+\frac{C_1\sqrt{\Delta_1}}{V_{\phi\xi}''}\right)
\tilde{D}_4t\right)e^{-(H_f + \sqrt{\Delta_1})t/2}\right\}.
\end{array}
\end{equation*}

The sufficient conditions for the asymptotic stability of the
considered fixed point are
\begin{equation}
H_f>0, \qquad \frac{V_{\xi\xi}}{C_2}+\frac{V_{\phi\phi}}{C_1}>0.
\label{V2fCond2}
\end{equation}
\item In the case
\begin{equation*}
{V_{\phi\xi}''}^2=\frac{C_1C_2}{16}\left(9H_f^2-\frac{4V_{\xi\xi}''}{C_2}\right)
\left(9H_f^2-\frac{4V_{\phi\phi}''}{C_1}\right)={} -
\frac{C_1C_2}{4}\left(\frac{V_{\xi\xi}''}{C_2}-\frac{V_{\phi\phi}''}{C_1}\right)^2
\end{equation*}
it is easy to check that the equality $\Delta_1= 0$ is valid
automatically and, therefore,
 the solution of (\ref{x})--(\ref{y}) is

\begin{equation}
\begin{array} {c}
\displaystyle{\phi_1(t)}=\left(\tilde{D}_1+\tilde{D}_2t+\tilde{D}_3t^2+\tilde{D}_4t^3\right)e^{-\frac{3}{2}H_ft},
\end{array}
\label{af}\
\end{equation}
\begin{equation}
\begin{array} {c}
\displaystyle{\xi_1(t)}=\frac{C_1}{2V_{\phi\xi}''}\left\{\tilde{D_1}
\left(\frac{V_{\xi\xi}''}{C_2}-\frac{V_{\phi\phi}''}{C_1}\right)+
\tilde{D_2}\left(\frac{V_{\xi\xi}''}{C_2}-\frac{V_{\phi\phi}''}{C_1}\right)t\right.+\\
\displaystyle\left.\tilde{D_3}\left[\left(\frac{V_{\xi\xi}''}{C_2}
-\frac{V_{\phi\phi}''}{C_1}\right)t^2-4\right]+\tilde{D_4}\left[\left(\frac{V_{\xi\xi}''}{C_2}
-\frac{V_{\phi\phi}''}{C_1}\right)t^3-12t\right]\right\}e^{-\frac{3}{2}H_ft}.
\end{array}
\label{ag}
\end{equation}

For the asymptotic stability of the considered fixed point it is
sufficient that the inequality $H_f>0$ hold.

\end{enumerate}

So, we obtained that in the case $V_{\phi\xi}\neq0$ the general
sufficient conditions of asymptotical stability of the fixed point
$y_f=(H_f,\phi_f,\psi_f)$ of the system of equations~(\ref{SYSTEM2})
are as follows
\begin{equation}
H_f>0, \qquad \frac{V_{\xi\xi}''}{C_2}+\frac{V_{\phi\phi}''}{C_1}>0,
\qquad
\frac{V_{\xi\xi}''V_{\phi\phi}''}{C_1C_2}>\frac{{V_{\phi\xi}''}^2}{C_1C_2}.
\label{V2fCondVVV}
\end{equation}

It is easy to see that the conditions (\ref{V2fCondVFKVFK}) obtained in
the case $V_{\phi\xi}''=0$ are equivalent to the conditions
(\ref{V2fCondVVV}), if the equality $V_{\phi\xi}''=0$ is substituted in
them.

So, in the general case including all the above particular cases with
definite relations between the parameters, the sufficient conditions
for asymptotic stability of the fixed point  $y_f=(H_f,\phi_f,\psi_f)$
of the system of equations~(\ref{SYSTEM2}) are (\ref{V2fCondVVV}).

Let us add the cold dark matter (CDM) to our model. One-field models
with the CDM in the Bianchi I metric have been considered
in~\cite{ABJV09}. The generalization for two-field models is
straightforward, so we point only the most important steps. If one adds
into consideration the CDM energy density $\rho_m$, then
system~(\ref{SYSTEM2}) should be modified as follows:
\begin{equation}
\begin{array}{l}
\displaystyle \dot H={}-3H^2+8\pi G_N\left(V(\phi,\xi)+\rho_m\right),\\
\displaystyle \dot\phi=\psi,\\
\displaystyle \dot{\psi}={}-3H\psi-\frac{1}{C_1}\frac{\partial V}{\partial\phi},\\
\displaystyle \dot\xi=\zeta,\\
\displaystyle \dot{\zeta}={}-3H\zeta-\frac{1}{C_2}\frac{\partial
V}{\partial\xi}\\
\displaystyle\dot\rho_m={}-3H\rho_m.
\end{array}
 \label{eomcdm}
\end{equation}

Let us consider the possible fixed points of system (\ref{eomcdm}).
From the last equation of this system, it follows that  at the fixed
point we have either $H_f=0$ or $\rho_{mf}=0$. Substituting
(\ref{m})--(\ref{q}) and
\begin{equation}
\rho_{m}(t)=\rho_{mf}+\varepsilon\tilde{\rho}_{m}(t)+{\cal
O}(\varepsilon^2),
\end{equation}
into the system (\ref{eomcdm}), we obtain the following system to
first order in $\varepsilon$:

\begin{eqnarray}
\displaystyle \dot h(t)&=&{}-6H_fh(t)+8\pi
G_N\tilde{\rho}_m(t),\label{sv1}
\\
\dot{\tilde{\rho}}_m(t)&=&{}-3H_f\tilde{\rho}_m(t)-3\rho_{mf}h(t),\\
\displaystyle \dot\phi_1(t)&=&\displaystyle \psi_1(t),\label{sv2} \\
\displaystyle \dot\psi_1(t)&=&\displaystyle {}-3H_f\psi_1(t)-
\frac{1}{C_1}\left(V''_{\phi\phi} \phi_1(t) +V''_{\phi\xi}
\xi_1(t)\right), \\
\displaystyle \dot\xi_1(t)&=&\displaystyle \zeta_1(t), \\
\displaystyle \dot\zeta_1(t)&=&\displaystyle {}-3H_f\zeta_1(t)-
\frac{1}{C_2}\left(V''_{\xi\phi} \phi_1(t) +V''_{\xi\xi}
\xi_1(t)\right),
\end{eqnarray}

It is easy to see that the last four equations of this system coincide
with the equations (\ref{s})--(\ref{v}). Therefore, the case $H_f=0$
can not be analysed with the Lyapunov theorem. Let us prove that
conditions (\ref{V2fCondVVV}) are sufficient for the stability of fixed
points for models with the CDM. First of all, it follows from $H_f\neq
0$  that $\rho_{mf}=0$. Solving equations (\ref{sv1})--(\ref{sv2}), we
obtain
\begin{equation}
\tilde{\rho}_m(t)=b_1e^{-3H_ft},\qquad
h(t)=b_0e^{-6H_ft}+\frac{b_1}{3H_f}e^{-3H_ft},
\end{equation}
where $b_0$ and $b_1$ are arbitrary constants.

We come to the conclusion  that if conditions (\ref{V2fCondVVV}) hold,
then a solution of system of equations (\ref{eomcdm}) is stable. In
other words if conditions (\ref{V2fCondVVV}) are satisfied, then the
solution, which is stable in the model without the CDM, is stable with
respect to the CDM energy density fluctuations as well.

Let us compare the obtained results with the known results.
In~\cite{Lazkoz} the stability of fixed points in quintom models
($C_1=-1$, $C_2=1$) with an arbitrary potential has been considered in
the FRW metric. The authors have presented the Einstein equations in
the form of a dynamical system having introduced the Hubble-normalized
variables~\cite{Wainwright_Lim}
\begin{equation} x_\phi=\frac{\dot\phi}{\sqrt{6}H},\qquad
x_\xi=\frac{\dot\xi}{\sqrt{6}H},\qquad y=\frac{\sqrt V}{\sqrt 3
H},\label{vars}
\end{equation}
and considered a system of ordinary differential equations with
differentiation with respect to the new variable $\tau=\log a^3$
instead of $t$. In our paper we have demonstrated that such change of
variables is not necessary, because the stability conditions can be
easily found in the initial variables.

Proceeding from some physical considerations, the authors
of~\cite{Lazkoz} assumed that $H_f>0$, we proved that the condition
$H_f<0$ is sufficient for instability of the fixed point. Note that the
case $H_f=0$, or equivalently $V(\phi_f,\xi_f)=0$, has not been
analysed both in~\cite{Lazkoz} and in our paper. In the case $H_f>0$ we
and the authors of~\cite{Lazkoz} obtained the same conditions on the
potential, which guarantee the stability of the fixed point. In our
paper we have proved that the obtained conditions are sufficient for
the stability not only in the FRW metric, but also in the Bianchi I
metric.

\section{Construction of stable solutions via the superpotential method}

\subsection{The superpotential for two-field models}

Let us consider the superpotential method~\cite{DeWolfe} (see
also~\cite{Superpotential,AKV}) for two-field models. We consider the
FRW metric and assume that the Hubble parameter $H(t)$ is a function
(superpotential) of $\phi(t)$ and $\xi(t)$:
\begin{equation}H(t)=W(\phi(t),\xi(t)),
\end{equation}
and that functions $\phi(t)$ and $\xi(t)$ are solutions of the
following system of two ordinary differential equations
\begin{equation}
\label{FG}
    \dot\phi=F(\phi,\xi), \qquad \dot\xi=G(\phi,\xi).
\end{equation}

Therefore,
\begin{equation}
    \ddot\phi=\frac{\partial F}{\partial \phi}F+\frac{\partial F}{\partial
    \xi}G,
\qquad
    \ddot\xi=\frac{\partial G}{\partial \phi}F+\frac{\partial G}{\partial
    \xi}G,
\end{equation}
and we can rewrite the Friedmann equations as follows:
\begin{equation}
\label{equH2W} 3W^2=8\pi
G_N\left(\frac{C_1}{2}F^2+\frac{C_2}{2}G^2+V\right),
\end{equation}
\begin{equation}
\frac{\partial W}{\partial\phi}F+\frac{\partial
W}{\partial\xi}G={}-4\pi G_N\left(C_1F^2+C_2G^2\right),
\end{equation}
\begin{equation}
\frac{\partial F}{\partial \phi}F+\frac{\partial F}{\partial
    \xi}G+3WF+\frac{1}{C_1}\frac{\partial V}{\partial\phi}=0,
\end{equation}
\begin{equation}
\frac{\partial G}{\partial \phi}F+\frac{\partial G}{\partial
    \xi}G+3WG+\frac{1}{C_2}\frac{\partial V}{\partial\xi}=0.
\end{equation}

From (\ref{equH2W}) we obtain
\begin{equation}
6W\frac{\partial W}{\partial \phi}=8\pi G_N\left(C_1F\frac{\partial
F}{\partial \phi}+C_2G\frac{\partial G}{\partial \phi}+\frac{\partial
V}{\partial \phi}\right),
\end{equation}
therefore,
\begin{equation}
G\left(\frac{C_2}{C_1}\frac{\partial G}{\partial
\phi}-\frac{\partial F}{\partial
    \xi}\right)=3W\left(\frac{1}{4\pi G_NC_1}\frac{\partial
W}{\partial \phi}+F\right).
\end{equation}

Also, we have
\begin{equation}
F\left(\frac{C_1}{C_2}\frac{\partial F}{\partial
\xi}-\frac{\partial G}{\partial \phi}\right)=3W\left(\frac{1}{4\pi
G_NC_2}\frac{\partial W}{\partial \xi}+G\right).
\end{equation}

If the functions $F$ and $G$ satisfy the following condition:
\begin{equation}
\label{GF} \frac{\partial F}{\partial
    \xi}=\frac{C_2}{C_1}\frac{\partial G}{\partial \phi},
\end{equation}
then
    \begin{equation}
\frac{\partial W}{\partial \phi}={}-4\pi G_NC_1\,F,\qquad
\frac{\partial W}{\partial \xi}={}-4\pi G_NC_2\,G. \label{equW}
\end{equation}

Note that using condition (\ref{GF}) and formulas (\ref{equW}), one can
verify that
    \begin{equation}
\frac{\partial^2 W}{\partial \phi\partial\xi}=\frac{\partial^2
W}{\partial \xi\partial\phi}.
\end{equation}

Therefore, to obtain particular solutions of system
(\ref{a2})--(\ref{e2}) it suffices to require that the relations
\begin{equation}
\frac{\partial W}{\partial\phi}={}-4\pi G_NC_1\dot\phi, \qquad
\frac{\partial W}{\partial\xi}={}-4\pi G_NC_2\dot\xi
\label{deWolfe_method},
\end{equation}
\begin{equation}
V =\frac{3}{8 \pi G_N}W^2-\frac{1}{32\pi^2
G_N^2}\left(\frac{1}{C_1}\left(\frac{\partial W}{\partial
\phi}\right)^2+\frac{1}{C_2}\left(\frac{\partial W}{\partial
\xi}\right)^2\right) \label{deWolfe_potential}
\end{equation}
be satisfied.

\subsection{Stability conditions in the superpotential method}

Let us obtain conditions on the superpotential $W$ that are equivalent
to conditions (\ref{V2fCondVVV}) for the corresponding potential $V$.
At the fixed point $y_f=(H_f,\phi_f,\psi_f)$ we get
\begin{equation}
W_f\equiv W(\phi_f,\xi_f)=H_f,\qquad  W'_\phi\equiv \frac{\partial
W}{\partial \phi}(\phi_f,\xi_f)=0,\qquad W'_\xi\equiv\frac{\partial
W}{\partial \xi}(\phi_f,\xi_f)=0. \label{DW0}
\end{equation}

It is easy to see that from conditions (\ref{DW0}) it follows that
\begin{equation}
V'_\phi=0,\qquad V'_\xi=0 \label{DV0}
\end{equation}
and
\begin{equation}
\label{SDVWphi2}
V''_{\phi\phi}=\frac{1}{16C_1C_2\pi^2G_N^2}\left(12C_1C_2\pi G_N W_f
W''_{\phi\phi}-C_2{W''_{\phi\phi}}^2-C_1{W''_{\phi\xi}}^2\right),
\end{equation}
\begin{equation}
\label{SDVWxi2}
V''_{\xi\xi}=\frac{1}{16C_1C_2\pi^2G_N^2}\left(12C_1C_2\pi
G_NW_fW''_{\xi\xi}-C_1{W''_{\xi\xi}}^2-C_2{W''_{\phi\xi}}^2\right),
\end{equation}
\begin{equation}
\label{SDVWphixi}
V''_{\phi\xi}=\frac{1}{16C_1C_2\pi^2G_N^2}W''_{\phi\xi}\left(12C_1C_2\pi
G_NW_f-C_2W''_{\phi\phi}-C_1W''_{\xi\xi} \right).
\end{equation}

The condition $H_f>0$ can be rewritten as $W_f>0$.

If $W''_{\phi\xi}=0$, then $V''_{\phi\xi}=0$ and  conditions
(\ref{V2fCondVFKVFK}) are
\begin{equation}
\left(12\pi G_NC_1 W_f -{W''_{\phi\phi}}\right)W''_{\phi\phi}
>0,\qquad
\left(12\pi G_NC_2W_f-{W''_{\xi\xi}}\right)W''_{\xi\xi}>0.
\end{equation}

In the general case conditions (\ref{V2fCondVVV}) are written in the
following form
\begin{equation}
12C_1C_2\pi
G_N(C_2W''_{\phi\phi}+C_1W''_{\xi\xi})W_f>C_2^2\left(W''_{\phi\phi})^2
+2C_1C_2(W''_{\phi\xi}\right)^2+C_1^2(W''_{\xi\xi})^2,
\end{equation}
\begin{equation}
\begin{array}{l}
\left( 144 C_1 C_2 \pi^2 G_N^2W_f^2-12C_1C_2\pi
G_N(C_2W''_{\phi\phi}+C_1W''_{\xi\xi})W_f+W''_{\xi\xi}W''_{\phi\phi}-(W''_{\phi\xi})^2\right)\times\\[2.7mm]
\times\left(W''_{\xi\xi}W''_{\phi\phi}-(W''_{\phi\xi})^2\right)>0.
\end{array}
\end{equation}

In the case of one-field models the stability conditions on the
superpotential are considered in Appendix.

\section{String field theory inspired cosmological models}

\subsection{Quintom models with the sixth degree potential}

  The interest in cosmological models related to the
open string field theory~\cite{IA} is caused by the possibility to get
solutions describing transitions from a perturbed vacuum to the true
vacuum (so-called rolling solutions~\cite{IA_TMF}). After all massive
fields (or some of the lower massive fields) are integrated out by
means of equations of motion, the open string tachyon acquires a
potential with a nontrivial vacuum, corresponding to a minimum of the
energy . The dark energy model~\cite{IA} (see also~\cite{AKV,AJ,AKV2})
assumes that our Universe is a slowly decaying D3-brane and its
dynamics is described by the open string tachyon mode. For the
Neveu--Schwarz--Ramond (NSR) open fermionic string with the GSO$(-)$
sector~\cite{NPB} in a reasonable approximation, one gets the Mexican
hat potential for the tachyon field (see~\cite{SFT-review} for a
review).  Rolling of the tachyon from the unstable perturbative
extremum towards this minimum describes, according to the Sen
conjecture~\cite{SFT-review}, the transition of an unstable D-brane to
a true vacuum.  In fact one gets a nonlocal potential with a string
scale as a parameter of nonlocality. After a suitable field
redefinition the potential becomes local, meanwhile, the kinetic term
becomes non-local. This nonstandard kinetic term  has a so-called
phantomlike  behavior and can be approximated by a phantom kinetic
term. It has been found that the open string tachyon behavior is
effectively modelled by a scalar field with a negative kinetic
term~\cite{AJK}. The back reaction of the brane is determined by
dynamics of the closed string tachyon. This dynamic can be effectively
described by a local scalar field $\xi$ with an ordinary kinetic
term~\cite{Oh} and possibly a nonpolynomial self-action~\cite{BZ}. An
exact form of the open-closed tachyon interaction is not known. So,
following~\cite{AKV2}, we consider the simplest polynomial interaction.

In the papers~\cite{AKV2,Vernov06} quintom models ($C_1={}- C_2<0$)
with effective potentials $V(\phi,\xi)$ have been
considered\footnote{Note that quintom models naturally arise from
nonlocal cosmological models with quadratic
potential~\cite{AJV2,Mulruny,Vernov2010}.}. The form of these
potentials are assumed to be given from the SFT within the level
truncation scheme. We postulate that the potential is a polynomial.
Specifically, we assume that the potential $V(\phi,\xi)$ should be an
even sixth degree polynomial
\begin{equation}
\label{potenV}
 V(\phi,\xi)=\sum_{k=0}^{6}\sum_{j=0}^{6-k}c_{kj}\phi^k\xi^j,\qquad
 V(\phi,\xi)=V(-\phi,-\xi),
\end{equation}
therefore, if the sum $k + j$ is odd, then $c_{kj}=0$.

From the SFT we can also assume asymptotic conditions for
solutions~\cite{AKV2,Vernov06}. We assume that the phantom field
$\phi(t)$ smoothly rolls from the unstable perturbative vacuum
($\phi=0$) to a nonperturbative one, for example $\phi=A$, and stops
there. The field $\xi(t)$ corresponds to the closed string and is
expected to go asymptotically to zero in the infinite future. Namely,
we seek such a function $\phi(t)$ that $\phi(0)=0$ and it has a
non-zero asymptotic at $t\to +\infty$: $\phi(+\infty)=A$. The function
$\xi(t)$ should have zero asymptotic at $t\to +\infty$. In other words,
we analyse the stability of solutions, tending to a fixed point with
$\phi_f=A$ and $\xi_f=0$.

\subsection{Construction of stable solutions}

Let us assume that there exists a polynomial superpotential
$W(\phi,\xi)$, which determines potential (\ref{potenV}) with formula
(\ref{deWolfe_potential}). To construct an even sixth degree polynomial
potential $V(\phi,\xi)$ we should choose $W(\phi,\xi)$ as an odd third
degree polynomial.  Obviously, the suitable form of the superpotential
is as follows:
\begin{equation}
\label{Wstring}
  W_{3}(\phi,\xi)=4\pi G_N\bigl(a_{1,0}\phi+a_{3,0}\phi^3+a_{0,1}\xi+a_{0,3}\xi^3
  +a_{2,1}\phi^2\xi+a_{1,2}\phi\xi^2\bigr),
\end{equation}
where $a_{i,j}$ are constants. For the superpotential $W_{3}$ system
(\ref{deWolfe_method}) is as follows:
\begin{equation}
\label{ODEstring}
\begin{split}
    \dot\phi&=\frac{1}{C_2}\left(a_{1,0}+3a_{3,0}\phi^2+2a_{2,1}\phi\xi+a_{1,2}\xi^2\right),\\
    \dot\xi&={}-\frac{1}{C_2}\left(a_{0,1}+3a_{0,3}\xi^2+a_{2,1}\phi^2+2a_{1,2}\phi\xi\right).
\end{split}
\end{equation}

Using asymptotic conditions: $\phi(+\infty)=A$, $\xi(+\infty)=0$,
$\dot\phi(+\infty)=\dot\xi(+\infty)=0$ we obtain
\begin{equation}
    a_{1,0}={}-3a_{3,0}A^2, \qquad a_{0,1}={}-a_{2,1}A^2.
\end{equation}

So, we obtain the following system of equations:
\begin{equation}
\begin{split}
    \dot\phi&=\frac{1}{C_2}\left(3a_{3,0}(\phi^2-A^2)+2a_{2,1}\phi\xi+a_{1,2}\xi^2\right),\\
    \dot\xi&={}-\frac{1}{C_2}\left(a_{2,1}(\phi^2-A^2)+3a_{0,3}\xi^2+2a_{1,2}\phi\xi\right)
\end{split}
\label{equW2}
\end{equation}
and the corresponding superpotential:
\begin{equation}
\label{Wstring2}
  W_3(\phi,\xi)=4\pi G_N\bigl({}-3a_{3,0}A^2\phi+a_{3,0}\phi^3-a_{2,1}A^2\xi+a_{0,3}\xi^3
  +a_{2,1}\phi^2\xi+a_{1,2}\phi\xi^2\bigr).
\end{equation}
At the fixed point $\phi_f=A$, $\xi_f=0$
\begin{equation}
  W_f={}-8\pi G_Na_{3,0}A^3.
\end{equation}
So, the condition $W_f>0$ is equivalent to $a_{3,0}A<0$.  It is easy to
calculate, that
\begin{equation}
W_{\phi\phi}''=24 \pi G_N a_{3,0}A, \qquad W_{\xi\xi}''=8 \pi G_N
a_{1,2}A, \qquad W_{\phi\xi}''=8 \pi G_N a_{2,1}A.
\end{equation}

Using (\ref{SDVWphi2})--(\ref{SDVWphixi}), we obtain:
\begin{equation}
V''_{\phi\phi}=\frac{4A^2}{C_2}\left(9(1-4\pi
G_NC_2A^2)a_{3,0}^2-a_{2,1}^2\right),
\end{equation}
\begin{equation}
V''_{\xi\xi}=\frac{4A^2}{C_2}\left(a_{2,1}^2-a_{1,2}^2-12\pi
G_NC_2A^2a_{3,0}a_{1,2}\right),
\end{equation}
\begin{equation}
V''_{\phi\xi}=\frac{4A^2a_{2,1}}{C_2}\left(3(1-4\pi
G_NC_2A^2)a_{3,0}-a_{1,2}\right).
\end{equation}

If $a_{2,1}=0$, then $V''_{\phi\xi}=0$ and the sufficient
conditions for the stability are
\begin{equation}
  a_{3,0}A<0, \qquad  4\pi G_NC_2A^2>1, \qquad a_{1,2}(a_{1,2}+12\pi G_NC_2 a_{3,0}A^2)<0.
  \label{StabcondWSTF1}
\end{equation}

If $a_{2,1}\neq 0$, then conditions (\ref{V2fCondVVV}) are equivalent
to
\begin{equation}
 \begin{split}
&  a_{3,0}A<0,\\
&2a_{2,1}^2{}-a_{1,2}^2-12\pi G_N C_2A^2a_{3,0}a_{1,2}+9(4\pi
G_NC_2A^2-1)a_{3,0}^2>0, \\
&\left[3a_{3,0}a_{1,2}-a_{2,1}^2\right]\left[3(4\pi
G_NA^2C_2-1)a_{3,0}a_{1,2}-36\pi G_NA^2C_2(4\pi
G_NA^2C_2-1)a_{3,0}^2+a_{2,1}^2\right]<0.
\end{split}
\label{w3cond}
\end{equation}

\subsection{Examples of stable solutions}

The case of superpotential $W(\phi,\xi)$ with $a_{2,1}=0$ and
$a_{0,3}=0$ has been considered in~\cite{AKV2}. In this case the system
(\ref{equW2}) has the following form
\begin{equation}
   \phi={}-\frac{C_2}{2a_{1,2}}\frac{\dot\xi}{\xi},
\end{equation}
\begin{equation}
    \ddot\xi=\frac{(2a_{1,2}-3a_{3,0})}{2a_{1,2}\xi}\dot\xi^2
    +\frac{2a_{1,2}}{C_2^2}\xi\left(3a_{3,0}A^2-a_{1,2}\xi^2\right).
\label{equxi}
\end{equation}
Equation (\ref{equxi}) can be integrated in quadratures:
\begin{equation}
\int\!\frac{\sqrt{\xi^{3 B-2}(3B+2)}C_2}{\sqrt{(12   B A^2+8 A^2-4
\xi^2)\xi^{3 B}a_{1,2}^2+(3 B+2) C_2^2D_1}}\:d\xi=\pm(t-t_0),
\end{equation}
where $D_1$ and $t_0$ are arbitrary constants and $B=a_{3,0}/a_{1,2}$.
At $B=-1/3$, $a_{2,1} = 0$, and $a_{0,3} = 0$, the general solution of
system (\ref{equW2}) can be obtained in explicit form~\cite{Vernov06}:
\begin{equation}
\phi_s(t)=\frac{A\left(C_2^2e^{4a_{1,2}At/C_2}-64a_{1,2}^4C_2^2A^2D_1^2-4a_{1,2}^2A^2D_2^2\right)}{\left(C_2e^{2
a_{1,2} A t/C_2}-2 D_2 a_{1,2} A\right)^2+64 D_1^2 a_{1,2}^4 C_2^2
A^2},
\end{equation}
\begin{equation}
\xi_s(t)=\frac{16 D_1 C_2^2 a_{1,2}^2 A^2e^{2 a_{1,2} A
t/C_2}}{\left(C_2e^{2 a_{1,2} A t/C_2}-2 D_2 a_{1,2} A\right)^2+64
D_1^2 a_{1,2}^4 C_2^2 A^2}.
\end{equation}

Let us analyse the stability of the exact solution. One can see
that $\phi_s(t)$ and $\xi_s(t)$ are continuous functions, which
tend to a fixed point at $t\rightarrow\infty$. Therefore, the
obtained exact solution is attractive if and only if the fixed
point is asymptotically stable.  At $a_{3,0}={}-a_{1,2}/3$ we
obtain that three stability conditions (\ref{StabcondWSTF1})
transform into two independent conditions
\begin{equation}
a_{1,2}A>0,\qquad C_2>\frac{1}{4\pi G_NA^2}.
\end{equation}
From these conditions it follows that $a_{1,2}A/C_2>0$.

Let us check the stability of solutions, obtained in~\cite{Vernov06}.
In~\cite{Vernov06} the author has considered the quintom model with the
following energy density:
\begin{equation}
\rho=\frac{8\pi G_N}{m_p^2}\left({}-\frac12\dot\phi^2
+\frac12\dot\xi^2+V_1\right),
\end{equation}
where
\begin{equation} m_p^2=\frac{g_o}{8\pi G_N M_s^2},
\end{equation}
$g_o$ is the open string coupling constant, $M_s$ is the string mass,
\begin{equation}
V_1=\frac{\omega^2
}{8A^2}\left(\left(A^2-\phi^2+\xi^2\right)^2-4\phi^2\xi^2+
\frac{1}{6m_p^2}\phi^2\left(3A^2-\phi^2+3\xi^2\right)^2\right),
\end{equation}
where $\omega$ is a nonzero constant.

Therefore, we obtain $C_2=M_s^2/g_o$, the potential
\begin{equation}
V_s=\frac{1}{8\pi G_Nm_p^2}V_1=\frac{M_s^2}{g_o}V_1
\end{equation}
corresponds to the superpotential
\begin{equation}
W_s=4\pi
G_N\omega\phi\left(A\left(\frac{1}{2}-\frac{\phi^2}{6A^2}\right)+
\frac{\xi^2}{2A}\right).
\end{equation}

This choice corresponds to
\begin{equation}
a_{1,2}=\frac{\omega}{2A},
\end{equation}
so the stability conditions are
\begin{equation}
\omega>0,\qquad 4\pi G_NA^2M_s^2>g_o.
\end{equation}

So, we come to the conclusion that the exact solutions, obtained
in~\cite{Vernov06} are stable for sufficiently large~$A$.

\section{Conclusion}

We have analysed the stability of isotropic solutions for
two-field models in the Bianchi I metric. Using the Lyapunov
theorem we have found sufficient conditions of stability of
kink-type and lump-type isotropic solutions for two-field models
in the Bianchi I metric.  The obtained results allow us to prove
that the exact solutions, found in string inspired phantom
models~\cite{AKV2,Vernov06}, are stable.

Our study  of the stability of isotropic solutions for quintom models
in the Bianchi I metric shows that the NEC is not a necessary condition
for classical stability of isotropic solutions. In this paper we have
shown that the models~\cite{AKV2,Vernov06} have stable isotropic
solutions and that large anisotropy does not appear in these models.
It means that considered models are acceptable, because they do not violate
limits on anisotropic models, obtained from the observations~\cite{Barrow,Bernui:2005pz}.

We also have presented the algorithm for construction of kink-type and
lump-type isotropic exact stable solutions via the superpotential
method. In particular we have formulated the stability conditions in
terms of superpotential.

This work is supported in part by state contract of Russian Federal
Agency for Science and Innovations 02.740.11.5057 and by RFBR grant
08-01-00798. I.A. is supported in part by grant of the Program for
Supporting Leading Scientific Schools NSh-795.2008.1. S.V. is supported
in part by grant of the Program for Supporting Leading Scientific
Schools NSh-1456.2008.2.

\section*{Appendix. Stability conditions on superpotential in a one-field cosmological model}

The main goal of our paper is to consider stable solutions in two-field
models. At the same time it is convenient to remind the superpotential
method for a cosmological model with one scalar field $\tilde{\phi}$,
which is described with the action
\begin{equation}
 S=\int d^4x \sqrt{-g}\left(\frac{R}{16\pi G_N}-
 \frac{\tilde{C}}{2}g^{\mu\nu}\partial_{\mu}\tilde{\phi}\partial_{\nu}\tilde{\phi}
-\tilde{V}(\tilde{\phi})\right), \label{action_1}
\end{equation}
where the potential $\tilde{V}(\tilde{\phi})$ is a twice continuously
differentiable function and $\tilde{C}$ is a nonzero real constant.

It has been shown in~\cite{ABJV09} that to find sufficient conditions
for the stability of the isotropic fixed point in the Bianchi I metric
one can consider the spatially flat Friedmann--Robertson--Worker
Universe with
\begin{equation*}
ds^2={}-dt^2+\tilde{a}^2(t)\left(dx_1^2+dx_2^2+dx_3^2\right),
\end{equation*}
where $\tilde{a}(t)$ is the scale factor. In the FRW metric the field
$\tilde{\phi}$ depends only on time.

The Friedmann equations can be written in the following form:
\begin{equation}
\dot{\tilde{H}}={}-4\pi
G_N\tilde{C}\left(\dot{\tilde{\phi}}\right)^2,\qquad 3\tilde{H}^2=8\pi
G_N\left(\frac{\tilde{C}}{2}\left(\dot{\tilde{\phi}}\right)^2+\tilde{V}(\tilde{\phi})\right).\label{eom12}
\end{equation}

In~\cite{ABJV09} it has been proven that the fixed point
$\tilde{y}_f=(\tilde{H}_f,\tilde{\phi}_f)$ is asymptotically stable
and, therefore, the exact solution $(\tilde{\phi}(t),\tilde{H}(t))$,
which tends to this fixed point, is attractive if:
\begin{equation}
\frac{\tilde{V}''(\tilde{\phi}_f)}{\tilde{C}}>0\quad \mbox{and}\quad
\tilde{H}_f>0,\label{Stabcond}
\end{equation}
in this subsection a prime denotes a derivative with respect to
$\tilde{\phi}$.

System of equations (\ref{eom12}) with a polynomial potential
$\tilde{V}(\phi)$ is not integrable. At the same time it is possible to
construct the potential $\tilde{V}(\tilde{\phi})$ and to find
$\tilde{H}(t)$ if $\tilde{\phi}(t)$ is given explicitly.

Following~\cite{DeWolfe}, we assume, that $\tilde{H}(t)$ is a function
of $\tilde{\phi}(t)$, called superpotential (for details of the
Hamilton--Jacobi formulation of the Friedmann equations and the
superpotential method see
also~\cite{Superpotential,AKV,AKV2,Vernov06}), that is
$\tilde{H}(t)=\tilde{W}(\tilde{\phi}(t))$. Using equality
$\dot{\tilde{H}}=\tilde{W}^{\prime}\dot{\tilde{\phi}}$, where
$\tilde{W}^{\prime}\equiv\frac{\partial
\tilde{W}}{\partial\tilde{\phi}}$, we obtain from system (\ref{eom12}):
\begin{eqnarray}
\dot{\tilde{\phi}}&=&{}-\frac{1}{4\pi G_N\tilde{C}}
\tilde{W}^{\prime},\label{eom1W}\\
\tilde{V}&=&\frac{3}{8\pi
G_N}\tilde{W}^2-\frac{1}{32\pi^2G_N^2\tilde{C}}\left(\tilde{W}'\right)^2.
\label{eom2W}
\end{eqnarray}

The superpotential method is to choose $\tilde{W}(\tilde{\phi})$ in
such form that both $\tilde{\phi}(t)$ and $\tilde{V}(\tilde{\phi})$
have required properties. Equation (\ref{eom1W}) is always solvable at
least in quadratures. Formula (\ref{eom2W}) allows one to find the
potential $\tilde{V}$, provided the superpotential $\tilde{W}$ is
given.

Let $\tilde{\phi}(t)$ tend to a finite limit $\tilde{\phi}_f$ at
$t\rightarrow +\infty$. We assumed that $\tilde{V}(\tilde{\phi})$ is a
twice continuously differentiable function, therefore,
$\tilde{V}(\tilde{\phi}_f)$ is finite and
$\tilde{H}_f=\tilde{W}(\tilde{\phi}_f)$ is finite as well. So, system
(\ref{eom12}) has the fixed point
$\tilde{y}_f=(\tilde{H}_f,\tilde{\phi}_f)$. It is easy to see that
\begin{equation}
\tilde{V}'(\tilde{\phi}_f)=0, \qquad \tilde{H}_f^2=\frac{8}{3}\pi
G_N\left(\tilde{V}(\tilde{\phi}_f)\right).
\end{equation}

From (\ref{eom2W}) we get  the condition
\begin{equation}
\tilde{V}'(\tilde{\phi}_f)=\frac{\tilde{W}'(\tilde{\phi}_f)}{16\pi^2G_N^2\tilde{C}}
\left(12\pi G_N
\tilde{C}\tilde{W}(\tilde{\phi}_f)-\tilde{W}''(\tilde{\phi}_f)\right)=0.
\end{equation}

If $\tilde{W}'(\tilde{\phi}_f^{\vphantom{27}})\neq 0$, then from
(\ref{eom1W}) it follows that $\tilde{\phi}_f$ is not a fixed point,
so, we analyse only the case
$\tilde{W}'(\tilde{\phi}_f^{\vphantom{27}})=0$ and obtain that
\begin{equation}
\tilde{V}''(\tilde{\phi}_f)
=\frac{\tilde{W}''(\tilde{\phi}_f)}{16\pi^2G_N^2\tilde{C}} \left(12\pi
G_N
\tilde{C}\tilde{W}(\tilde{\phi}_f)-\tilde{W}''(\tilde{\phi}_f)\right).
\end{equation}

Thus, we come to the conclusion that to construct a stable kink-type
solution one should find such $\tilde{W}(\tilde{\phi})$ that
$\tilde{\phi}(t)$ tends to a fixed point $\phi_f$ and the following
conditions are satisfied
\begin{equation}\label{StabcondW1}
\tilde{W}''(\tilde{\phi}_f^{\vphantom{27}})\left(12\pi G_N
\tilde{C}\tilde{W}(\tilde{\phi}_f)-\tilde{W}''(\tilde{\phi}_f)\right)>0\quad\mbox{and}\quad
\tilde{H}_f=\tilde{W}(\tilde{\phi}_f)>0.
\end{equation}

Note that the obtained conditions are sufficient for the stability of
the obtained isotropic solution in the Bianchi I metric as well as with
respect to small fluctuations of the CDM energy density~\cite{ABJV09}.

\end{document}